# Miniature shock tube for laser driven shocks

Michel Busquet (1), Patrice Barroso (2), Thierry Melse (2), Daniel Bauduin (2)

[1] LERMA, UMR 8112, Observatoire de Paris, CNRS and UPMC, 5 place J. Janssen, 92195 Meudon, France

[2] GEPI, UMR 8111, Observatoire de Paris, CNRS and Université Diderot, 77 Avenue Denfert Rochereau, 75014 Paris, France

**Corresponding author** : Michel Busquet, Observatoire de Paris, LERMA, 5 Place Jules Janssen, 92195 Meudon France, +33 1 450 774 16

mail : michel.busquet@obspm.fr

**Abstract**

We describe in this paper the design of a miniature shock tube (smaller than 1 cm3) that can be placed in a vacuum vessel and allows transverse optical probing and longitudinal backside XUV emission spectroscopy. Typical application is the study of laser launched radiative shocks, in the framework of what is called "laboratory Astrophysics".

## 1 –Introduction

A new domain of experimental physics has been active for a couple of years, named "laboratory astrophysics". The underlying idea is that phenomena observed in Astrophysics can generally not be completely measured, and can not be controlled nor done several times. So one need to scale down these phenomena to laboratory scale, and be able to control conditions, to redo the experiments as many times as needed, and to measure various parameters or to study a separate part of the whole phenomenon. Among the scalable phenomena of interest, one can found hydrodynamic instabilities, planet core Equation of State, and Radiative Shocks which is the matter of the present paper. Radiative Shocks [1] are super-critical shocks, where the emitted radiation is large enough to launch a radiation driven ionization wave (called radiative precursor) in front of the density jump and to cool down the post-shock region. In the Universe, such radiative shocks are found in accretion shock of matter falling on Young Stellar Objects, in proto-stellar disks, T Tauri stars or binary systems, as well as in outflows into circumstellar medium from Supernovae and Supernovae Remnants.

Scaling laws have been derived by different authors [1], [2]. They end in experiments on laser facilities where a miniature shock tube, filled with a low pressure high atomic number gas, is placed inside an enclosure pumped down to a low pressure ($P<10^{-4}$ torr) so the laser at high irradiance can propagate until the "piston" of the miniature shock tube. The challenge is to control dimensions to the required values *but* to allow probing of the shock in these tiny assay. Various means to diagnose the created shock and its radiative precursor have been used by different teams: image of the self-emission of a solid foil impacted by the precursor,[3] interferometric measurement,[4],[5] x-ray radiography [6] and Thomson scattering.[7] During the experiment we performed on the PALS facility [8], we used interferometric measurement of the electronic density in the precursor for a large traveling distance (up to 6 mm), thanks to the targets we specially designed for this experiment and that we describe in this paper. We first present in the next section the general setup of the experiment and the requirements for the target. In Sec.3 we describe in detail these targets, a sub-centimetric assembly of a miniature shock tube, the flyer (or "piston"), the holding base and the pump and fill circuit. The piston is to be launched by irradiation from a high-power laser (0.5 TW). We also describe a variant of these targets designed for backside spectroscopy that minimizes the perturbations brought by the joint of the shock tube to the base. This variant has allowed us to demonstrate the feasibility of XUV spectroscopy of radiative shocks.

2– Experimental setup and shock tube requirements

Target requirements are given by the experimental setup (Fig.1). The radiative shock is created in a miniature shock tube, closed at one end by a thin 10 microns plastic foil, to be acting as the piston. A powerful laser is focused on the plastic foil, smoothed with a Phase Zone Plate [9] that gives a focal spot of approximatively 550 microns FWHM, with 90% of energy in a diameter of 700 microns. Thus a 0.7 mm width of the canal is well suited. Larger focal spot would mean lower intensity, and lower shock velocity. We used the frequency tripled laser of the PALS facility [10] delivering ~ 150 J in 0.35 ns at $\lambda$=438 nm. The power of the laser pulse is half a TW and the laser intensity impacting the foil is around or above $10^{14}$ W/cm$^2$, which means that the target has to be placed in a vacuum chamber (below P=$10^{-4}$ torr) for the laser to reach the plastic foil. Laser ablation of the plastic accelerates the foil by rocket effect, ending in a piston velocity of a few 10 km/s when using a pressure of ~ 0.2 bar in the tube. The shock tube has to be filled with Xenon (or a mixture of air and Xenon) at a fraction of atmosphere (typically P=0.2 bar), as high Z material is required for a radiative shock to develop in front of the hydrodynamic shock created by a solid piston at such a velocity. The radiative precursor is probed by visible light interferometry, which allows measuring electron density [4], [8]. Also, we would like to measure the XUV emission in front of the shock, leading to close the back end of the tube with an XUV window. This backside XUV window, as well as the front side plastic foil has to withstand a differential pressure of up to 1 atm, in both directions :

- vacuum outside the tube (placed in the vacuum chamber) and up to 0.5 bar in the tube, when firing the laser
- or ambient pressure outside the tube, and zero pressure inside the tube when purging it from air before filling it with Xenon.

To observe long time behavior (up to 100 ns) a length of 6 mm is required as the average velocity of the precursor might reach 60 km/s. The target will be placed in one arm of a Mach-Zehnder interferometer, thus optical quality side windows covering the whole length of the tube is obviously required.

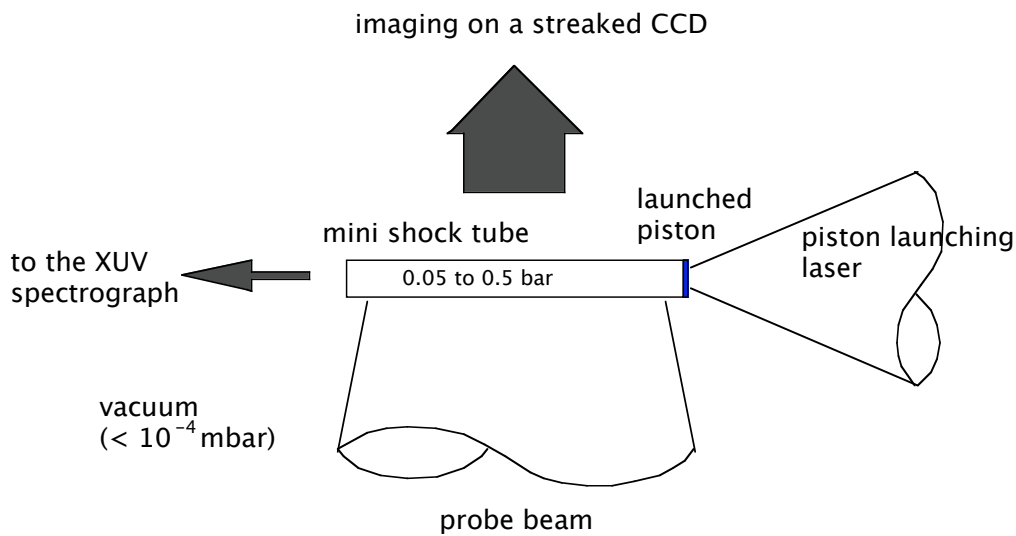

*Figure 1 schematic experimental layout*

In summary we want to make an assembly (or "target") with
- a tube of down to 0.7 mm transverse size
- and 6 mm length,
- with optical quality windows on 2 sides of the canal, which shall then be of square section,
- a 10 microns thin plastic foil at one end

- and an XUV window at the other,
- an appropriate pump and fill circuit with low perturbation on the shock propagation inside the tube
- and a convenient base plate for holding the target.
- material of the side walls should be controlled, as precursor propagation is sensitive to their albedo [11]

The experiment needs also that a 0.5 microns gold layer is deposited on the backside of the plastic foil to block x-rays coming from the laser interaction region.

A functional layout of the target is presented in Fig.2

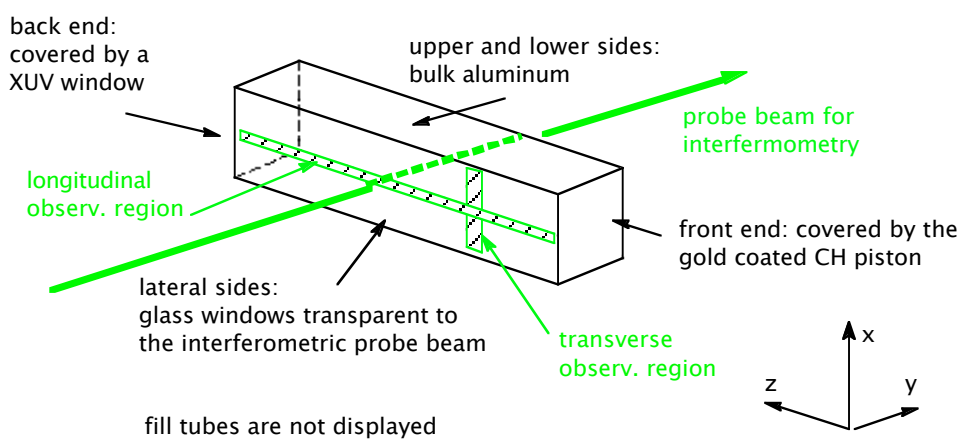

*Figure 2 layout of the shock tube shows its required functions*

In targets made for the only purpose of backside XUV spectroscopy, we want to further minimize perturbation to the shock propagation until the very end of the tube. An internal circular section can be used and will ease numerical simulation of the radiative precursor.

We successfully build such targets, packing in less than 1 cm3 assembly all of the required features. We now give details on the target design in the following sections.

**3- description of the targets for interferometric probing.**

The targets are made from micro-machined aluminium. They consist in a 2 mm thick, 12x12mm square base plate holding a two-prong "fork" (Fig.3). Optical quality BK7 glass windows are glued on the side of the prongs. The two prongs, that have a thickness of 0.7 mm, and the two glass windows glued on them are delimiting a 6 mm long 0.7 x 0.7 mm square channel, that will be our shock tube.

Wall material can be controlled by sputtering various components on both the aluminium piece and the inner side of the windows before assembling. For the experiment performed on the PALS facility in 2007 [8], we used either aluminium (10 nm coating) or gold (15 nm coating over a 2 nm coating of Chromium). Regarding the propagation of the low temperature (< 30 eV) radiative shock, these coating are thick enough to play the role of bulk material. For the interferometric probing, the transmission of the coated pair of windows is above 10%, and can be managed by unbalancing the intensity in the two arms of the Mach-Zehnder interferometer. The high-speed high-precision saw used to cut the glass windows produces square ends suitable to glue the plastic foil on top of them, with a vacuum tight connection.

A 0.6 mm hole is drilled in the center of the base plate for the two purpose of
- allowing back side XUV observation (it will be covered by a vacuum tight XUV transmitting window glued on the back of the base plate)
- creating a pump and fill circuit of the shock tube in connection with two holes drilled in the 2 mm thickness (Fig. 4) of the base plate, like secret passages in the walls of a medieval castle.

Two fill tubes (1mm in diameter) will be glued inside these holes, connected to a small 1cm$^3$ pressure ballast tank (Fig.5) that will allow the target to hold the required pressure for a couple of hours even if there were small leaks, or if there is gas releasing materials.

On one of the two prongs, we machine a set of grooves for magnification reference.

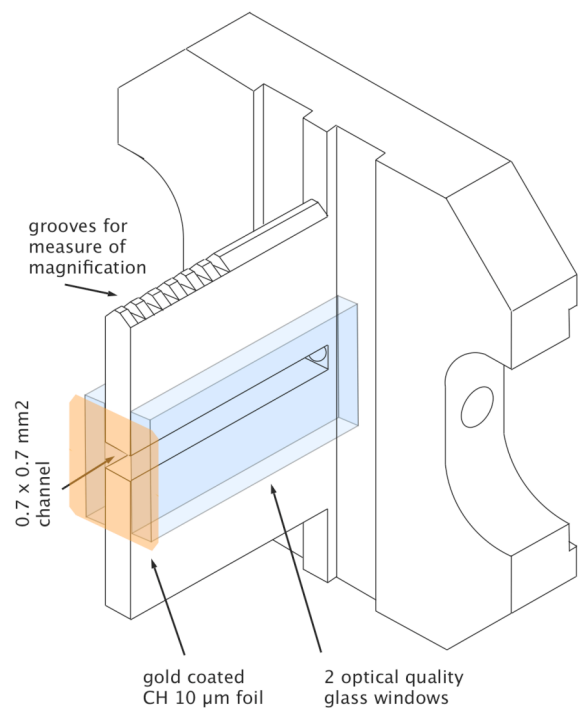

*Figure 3 - 3D drawing of the target, the two glass side windows and the gilded plastic foil. The XUV window is glued on the back of the base plate.*

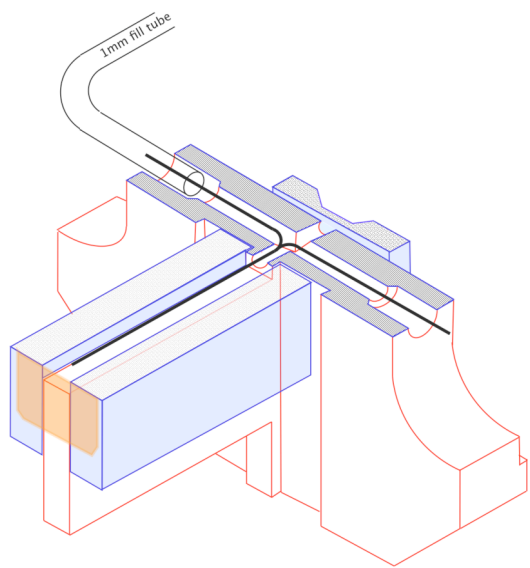

*Figure 4. Cutout of the target, mounted with the 4 windows (note the thinned wafer glued on the back), shows the pump and fill circuit (black lines).*

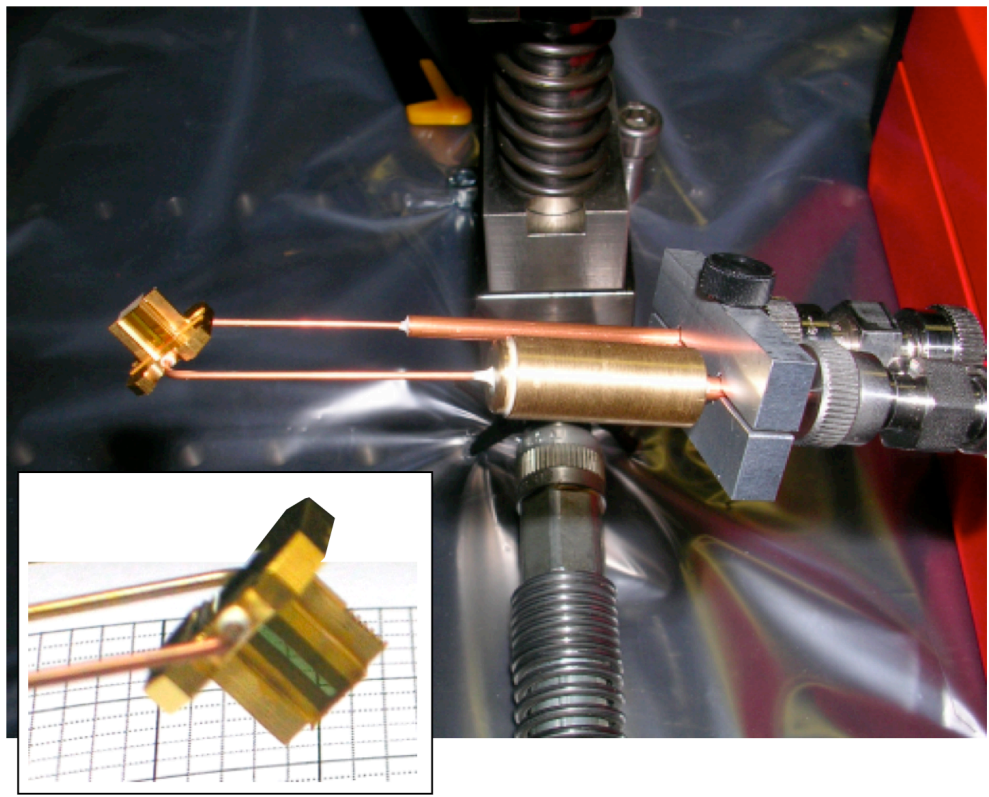

*Figure 5. Target is prepared on the pumping system. Note the 1 cm3 ballast tank. Inset shows details of the glued assemblage (courtesy of F.Thais).*

For a good XUV transmission, the backside XUV window can be made of a thin foil (0.1 to 0.2 μm) of either Aluminium, SiC or $Si_3N_4$. In the last 2 cases, a 0.5 mm thick wafer is used, chemically etched on a surface of 1 x 1 mm to the required thickness. For Al, we use grid-supported foils of the type classically used for scanning electron microscopy (SEM). As said previously, the channel is to be filled with 0.05-0.5 bar gas and has to withstand a 1 bar differential pressure in both directions. This directs the choice to polystyrene for making the front-end thin foil, not too much stiff (as it would break under pressure) and not too much elastic (as it would deflate when pumping the tube and inflate once placed inside the vacuum chamber). Polystyrene can also bear the high temperature encountered in the process of sputtering the thin gold overcoat required to stop x-rays originated from the laser interaction.

The target is to be placed in one arm of a Mach-Zehnder interferometer. The output of the interferometer is imaged on the entrance slit of a visible light streak camera. An example of time resolved interferogram of shock propagation in the tube is shown in Fig.6, demonstrating the possibilities of the target described in this paper.

The shock tube can even be made with a smaller section, down to 0.3 x 0.3 $mm^2$ with stepped prongs and staggered windows on the sides.

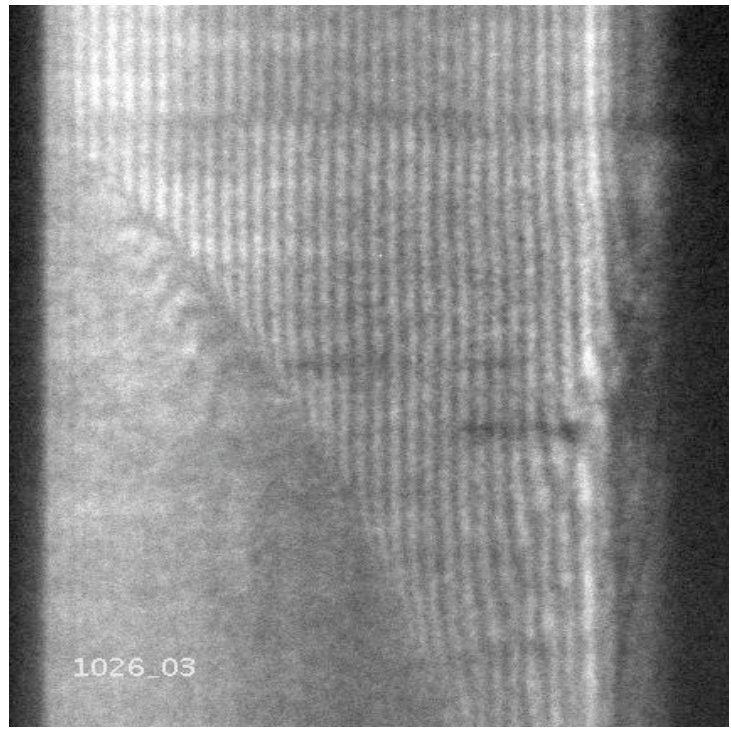

*Figure 6. Time resolved interferogram of the shock tube in action. The whole 6 mm length is imaged (horizontal axis) and the time observation last 100ns (vertical axis).*

**4- description of "blind" targets for backside spectroscopy.**

At the back end of the tube, pumping hole and base plate steps result in a complex geometry and will produce perturbation to the propagation of the radiative shock after the first 6 mm. Though it is not too big a problem for time resolved observation, the XUV spectrographic recordings we performed (through the central hole in the base plate) are time integrated, and would suffer from this perturbation. We thus design a different target with no lateral windows, In these "blind" targets, the channel is made from a cylindrical tube, encased in the base plate with flattened end to create the pump and fill circuit (Fig.7). A 0.1-0.2 mm spacing, between the end of the tube and the backside window, creates a damping expansion chamber, where the shock weakens before reaching the XUV window. Preliminary results obtained with a low sensitivity, slitless, XUV spectrograph are shown in Fig.8 with different conditions (wall material of Alumina or gold, gas fill of pure Xe or 50-50 mixture of Air and Xenon). The sharp cut-off observed at 17.5nm results from the L-edge of the Aluminum filter. Without slit in front of the spectrograph, the spectral resolution is given by the source size. A resolution of $\lambda/\Delta\lambda$ around 100 is obtained. Also, as we were working on the filter pack and the settings of the spectrograph, sensitivity varies from shots to shots. In the future, we shall use our high sensitivity, high spectral resolution and medium spatial resolution XUV spectrograph recently made for the purpose of direct measuring of wall albedo. [12]

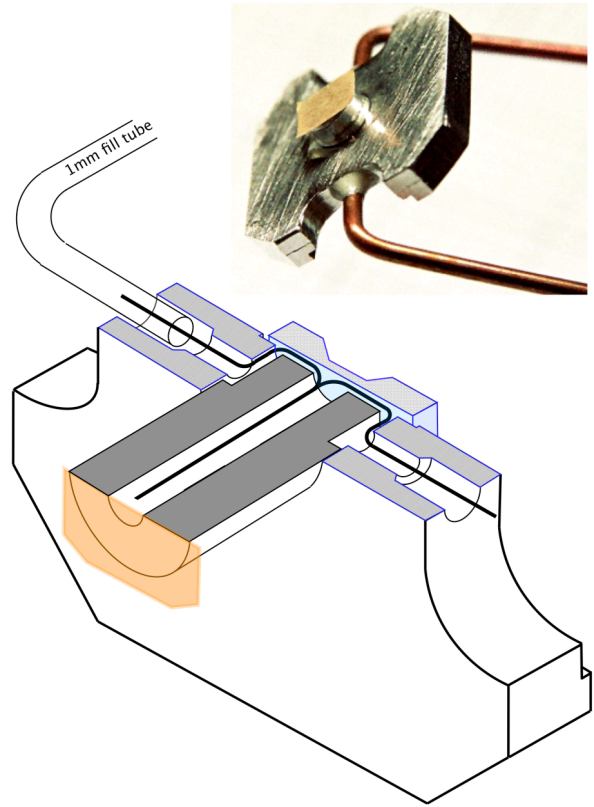

*Figure 7 - 3D cutout and photograph (courtesy of F.Thais) of a "blind target". The flattened end of the encased cylinder and the spacing between the window and the cylinder allows a pumping circuit (black thick line).*

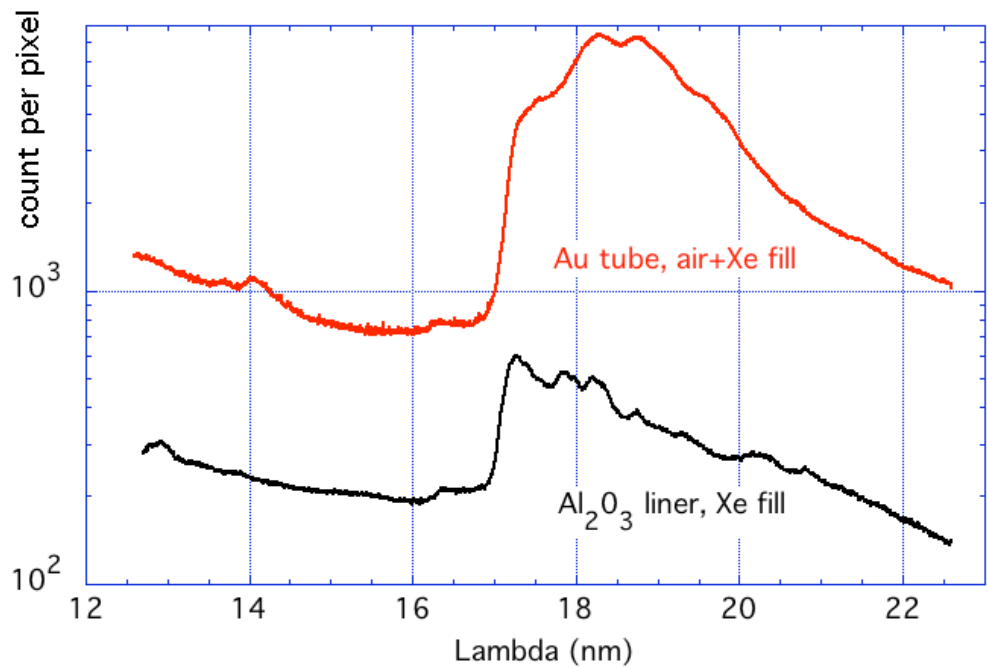

*Figure 8 - 2 recorded spectra with different conditions (wall material and gas fill). The spectral resolution $\lambda/\Delta\lambda$ around 100 is given by the source size as we used a slitless spectrograph. Note that the spectrograph filtering and setting was changed between shots, thus sensitivity varies from shots to shots.*

## 5 – TARGET OPERATION AND CYCLE TIME

The minimum time between two shots on such high energy (>100 J ) laser facility is between 20 and 40 minutes, due to laser operation requirements (cooling down of the laser components, laser chain alignment, capacitor bank charging, …). On the other hand opening and pumping down the vacuum chamber is around 10 minutes, and target alignment may also reach 10 minutes, depending on the viewing system implemented around the vacuum chamber. So the time required (a couple of minutes) to fill the target with the appropriate fill gas (or mixture) at the chosen pressure add no extra delay, as it can be done during the vacuum operating cycle. The piston and the windows are destroyed during the laser pulse or in the first few milliseconds after, but the metal casing can be analyzed "post-mortem" to check the quality of the laser pointing. Digital micro-machining of the socket and gluing the windows allow excellent reproducibility of the targets. On the PALS laser facility, aging of the Iodine active medium of the laser is now the main origin of the variability (contamination comes from glass cells, pumps and tubing). However, we have found that shots are fairly reproducible during the same day.

## 6 – CONCLUSIONS AND PERSPECTIVES

We successfully build targets, assembling in less than 1 cm3 all of the required features, and achieved miniature shock tubes of 6 mm length and down to 0.7 mm transverse dimension, with a design allowing to go down to 0.3 mm. Two essential features are 1) the possibility to place the miniature shock tube inside a vacuum chamber and 2) the optical quality of the tube allowing transverse probing in visible light. We are currently working on the feasibility of XUV transverse probing of modified targets.

Used on the PALS laser facility,[8] these targets have allowed to probe electron density in the shock tube for time up to 100 ns after the laser pulse. Recording of backside XUV emission with spectral resolution has also been demonstrated. Specific experiment to measure the wall albedo at a temperature of 10-30 eV is described elsewhere.[12] With laser facility of the 1 kJ class, and/or with a different design of the piston, shock velocities above 150 km/s should be within reach with the same target design.

## 7 – ACKNOWLEDGMENTS

We thank O. Madouri (LPN) and V.Petitbon (IPN) for providing us with the SiC windows and the polystyrene thin films. The interferometer was built in collaboration with Ouali Acef (Observatoire de Paris) and Frédéric Thais (CEA/DSM). We acknowledge the participation of Frédéric Thais, Chantal Stehlé, Guillaume Loisel, Norbert Champion, Michaela Kozlova, Tomas Mocek, Bedrich Rus and his team, during the experiment performed at PALS under the P.I.ship of Chantal Stehlé and Michel Busquet [13].

We obtained financial supports from the Access to Research Infrastructures activity in the Sixth Framework Programme of the EU (contract RII3-CT-2003-506350 LaserLab Europe), from the RTN JETSET (contract MRTN-CT-2004 005592), from CEA/DSM (Saclay), from Observatoire de Paris and from the French National Programme of Stellar Physics (PNPS), CNRS-PICS 4343. This research was partially supported by the Czech Ministry of Education, Youth and Sports (project LC528). We also acknowledge financial support from the OPTON LASER INTERNATIONAL Corporation.

## 8- REFERENCES

[1] Y.B.Zeldovich, Y.P. Raiser, (1966) Physics of Shock Waves and High Temperature Hydrodynamic Phenomena, New York: Academic Press; N. Kiselev Yu, I.V.Nemchinov, V.V.Shuvalov, Comput. Math. Phys., **31** (1991), 87-101

[2] D.Mihalas and B.Mihalas, "Foundations of radiation Hydrodynamics" (Oxford University, Oxford, 1984); G.C. Pomraning "The Equations of Radiation Hydrodynamics", (Pergamon, Oxford, 1973); C.Michaut, *et al*, Eur. Phys. J. D, **28**, 381 (2004)

[3] J.C.Bozier, G.Thiell, J.P.Lebreton, S.Azra, M.Decroisette, D.Schirmann, Phys.Rev.Lett. **57**, 1304 (1986)

[4] S.Bouquet, *et al,* Phys.Rev.Lett. **92**, 5001 (2004)

[5] M.Busquet, *et al*, High Energy Density Physics, **3**, 8 (2007); M.Gonzàlez, *et al*, Laser Part. Beams, **24**, 535 (2006)

[6] A.B. Reighard, *et al*, Phys. Plasmas **13**, 082901 (2006); A.B. Reighard, *et al*, Phys. Plasmas **14,** 056504 (2007)

[7] A.B. Reighard, D.R. Leibrandt, S. G. Glendinning, T. S. Perry, B. A. Remington, J. Greenough, J. Knauer and T. Boehly, S. Bouquet, L. Boireau, M. Koenig, T. Vinci, Rev. Sci. Inst. **77**, 10E503 (2006)

[8] M. Busquet*,* F. Thais, E. Audit, M. Gonzalez, J.Appl. Phys. **107**, 083302 (2010); C. Stehle, M. Gonzalez, M. Kozlov, B. Rus, T. Mocek, O. Acef, J.-P. Colombier, T. Lanz, N. Champion, K. Jakubzak, J. Polan, P. Barroso, D. Baudin, E. Audit, J. Dostal, and M. Stupka, "Experimental study of radiative shocks at PALS facility," Laser Part. Beams (in press)

[9] I.Ross, D.A.Pepler, C.N.Danson, Optics Comm., **116**, 55 (1995)

[10] K.Jungwirth, Laser Part. Beams, **23**, 177 (2005)

[11] S.Leygnac, L.Boireau, C.Michaut, T.Lanz, C.Stehlé, C.Clique, S.Bouquet, Phys. Plasmas, **13**, 113301 (2006); M. González, E.Audit, C.Stehlé, A&A, **497**, 27 (2009).

[12] M.Busquet, F.Thais, G.Geoffroy, D.Raffestin, Bull.Am.Phys.Soc., **53**, 301 (2008); M.Busquet, F.Thais, G.Geoffroy, D.Raffestin, Rev. Sci. Instr., in prep.

[13] http://www.pals.cas.cz/pals/projekty/proj1077.htm
http://www.pals.cas.cz/pals/pae104fr.htm
http://www.pals.cas.cz/pals/pae104ma.htm